\documentclass[fleqn,usenatbib]{mnras}

\usepackage{newtxtext,newtxmath}

\usepackage[T1]{fontenc}
\usepackage{ae,aecompl}

\usepackage{graphicx}	
\usepackage{amsmath}	

\title[Telluric correction of exoplanet transmission spectra]{Assessing telluric correction methods for Na detections with high-resolution exoplanet transmission spectroscopy}

\author[A. Langeveld et al.]{
Adam B. Langeveld,$^{1}$\thanks{E-mail: adam.langeveld@ast.cam.ac.uk}
Nikku Madhusudhan,$^{1}$\thanks{E-mail: nmadhu@ast.cam.ac.uk}
Samuel H. C. Cabot$^{2}$, 
Simon T. Hodgkin$^{1}$
\\
$^{1}$Institute of Astronomy, University of Cambridge, Madingley Road, Cambridge, CB3 0HA, UK\\
$^{2}$Yale University, 52 Hillhouse, New Haven, CT 06511, USA\\
}

\date{Accepted 2021 January 7. Received 2020 December 24; in original form 2020 August 27}

\pubyear{2021}

\begin{document}
\label{firstpage}
\pagerange{\pageref{firstpage}--\pageref{lastpage}}
\maketitle

\begin{abstract}
Using high-resolution ground-based transmission spectroscopy to probe exoplanetary atmospheres is difficult due to the inherent telluric contamination from absorption in Earth's atmosphere. A variety of methods have previously been used to remove telluric features in the optical regime and calculate the planetary transmission spectrum. In this paper we present and compare two such methods, specifically focusing on Na detections using high-resolution optical transmission spectra: (1) calculating the telluric absorption empirically based on the airmass, and (2) using a model of the Earth's transmission spectrum. We test these methods on the transmission spectrum of the hot Jupiter HD~189733~b using archival data obtained with the HARPS spectrograph during three transits. Using models for Centre-to-Limb Variation and the Rossiter-McLaughlin effect, spurious signals which are imprinted within the transmission spectrum are reduced. We find that correcting tellurics with an atmospheric model of the Earth is more robust and produces consistent results when applied to data from different nights with changing atmospheric conditions. We confirm the detection of sodium in the atmosphere of HD~189733~b, with doublet line contrasts of $-0.64 \pm 0.07~\%$ (D2) and $-0.53 \pm 0.07~\%$ (D1). The average line contrast  corresponds to an effective photosphere in the Na line located around $1.13~R_{\text{p}}$. We also confirm an overall blueshift of the line centroids corresponding to net atmospheric eastward winds with a speed of $1.8 \pm 1.2$~km~s$^{-1}$. Our study highlights the importance of accurate telluric removal for consistent and reliable characterisation of exoplanetary atmospheres using high-resolution transmission spectroscopy.
\end{abstract}



\section{Introduction}
\label{sec:introduction}

The era of characterising exoplanet atmospheres has accelerated greatly over the last two decades. With technological advancements in instrumentation, we are now able to monitor exoplanetary systems and characterise in detail both their bulk properties as well as their atmospheres. Numerous planets have been observed with both radial velocity and transit methods, allowing for extensive analyses and characterisation \citep[e.g.][]{fischer2014, sing2016, madhusudhan2019, pinhas2019, welbanks2019}. Observations of transiting systems can give an insight into the chemical properties of the planetary atmosphere. During one orbit there are two windows which can be used for such studies: the primary transit, when the planet passes in front of its host star as viewed from Earth, and the secondary eclipse when the planet is occulted by the star. In both instances, the planetary atmospheric spectrum can be deduced by measuring the change in observed flux outside of and during transit \citep{seager2000}, and absorption or emission due to chemical species may be seen. Such inferences can also be made if phase-resolved spectra are obtained at other parts of the orbit \citep{stevenson2014}.

The first observation of an exoplanet atmosphere was made by \citet{charbonneau2002}. Using the STIS spectrograph on the \textit{Hubble Space Telescope} (HST) with a resolution of $R = 5540$, absorption at 589~nm due to atmospheric sodium (Na \textsc{i}) was detected in the transmission spectrum of HD~209458~b. The first ground-based detection was made using the High Resolution Spectrograph (HRS) on the 9.2~m Hobby-Eberly telescope \citep{redfield2008}. With a resolution of $R \sim$~60,000, the sodium doublet was fully resolved and absorption in the atmosphere of HD~189733~b was detected. Shortly after, \citet{snellen2008} used the High Dispersion Spectrograph (HDS) ($R \sim$~45,000) on the 8~m Subaru telescope to confirm the presence of sodium in HD~209458~b. With a combination of low to medium-resolution space-based data and high-resolution ground-based data, there have been discoveries of almost 20 different chemical species in exoplanet atmospheres to date \citep{madhusudhan2019}. Among the most common are detections of the alkali species Na and K in hot Jupiters \citep[e.g.][]{charbonneau2002, redfield2008, sing2016, sedaghati2016, nikolov2016, wyttenbach2017, chen2018, casasayas-barris2018, jensen2018, deibert2019, seidel2019, hoeijmakers2019}.

Many challenges are presented when using ground-based facilities to acquire high-resolution spectra of transiting exoplanets. For methods which derive the planetary transmission spectrum by comparing the in-transit and out-of-transit stellar spectra, observing times must be chosen to ensure that an adequate out-of-transit baseline can be obtained. An insufficient out-of-transit baseline can adversely affect the quality of the transmission spectrum \citep{wyttenbach2015}. A significant problem arises from the contamination of spectra due to absorption in the Earth's atmosphere. These telluric lines can merge with stellar lines, or appear at a similar or higher strength as features in the planetary transmission spectrum \citep{redfield2008, snellen2008}. In the optical domain, the predominant sources of absorption are telluric water and oxygen. Therefore, data must be obtained on nights with low water vapour content and when the airmass is low to minimise the impact of the contamination.

Alongside the telluric removal process, it is essential to make further corrections to account for stellar reflex motion, systemic velocity, and the planetary radial velocity. Additionally, a planet which passes in front of a rotating star blocks different amounts of blueshifted and redshifted light. This causes the shape of lines in the integrated stellar spectrum to change as the transit progresses, and is known as the Rossiter-McLaughlin (RM) effect \citep{rossiter1924, mclaughlin1924, queloz2000, triaud2018}. The spectral line shape is also affected by Centre-to-Limb Variation (CLV) which describes the brightness of the stellar disc as a function of limb angle \citep{czesla2015, yan2017}. Together with telluric contamination, these effects may imprint spurious signals within the planetary transmission spectrum and must be corrected for to prevent false identifications of chemical species. A recent study has shown that absorption features in the transmission spectrum of HD~209458~b can be explained purely by the induced CLV and RM signals \citep{casasayas-barris2020}, highlighting the importance of correcting for these effects. In the near future, newly developed high-resolution spectrographs will be used to target rocky planets and super-Earths. It is therefore crucial to be able to accurately tackle the many problems associated with ground-based observations in order to distinguish spectral features of these planets. Further characterisation will be possible when these results are used in combination with upcoming \textit{James Webb Space Telescop}e (JWST) data.

Several methods have been used in previous work to remove telluric features from high-resolution spectra. For the first detection of the atmosphere of HD~189733~b from a ground-based facility \citep{redfield2008}, the spectrum of a telluric standard star was recorded immediately after the observations. A spectrum of absorption in the Earth's atmosphere was recovered by using a stellar template to remove the weak stellar lines of the rapidly rotating hot B-star. This was also employed in subsequent studies \citep{jensen2011, jensen2012}. Other datasets have been corrected by assuming a linear relationship between airmass and telluric line strength, and deriving a telluric spectrum to remove variation throughout the night \citep{vidal-madjar2010, astudillo2013, wyttenbach2015}. More recently, studies have used models of Earth's atmosphere to correct for the telluric spectrum, e.g. \texttt{TERRASPEC} \citep{lockwood2014}, \texttt{molecfit} \citep{smette2015, kausch2015, allart2017}, and other custom-built codes \citep[e.g.][]{yan2015, casasayas-barris2017}. Contamination can also be corrected by removing the time-dependent variation of flux at each wavelength element using linear regression \citep[e.g.][]{snellen2010, brogi2012}, or by removing systematic trends with singular value decompositions (SVDs) or principal component analysis (PCA) \citep[e.g.][]{dekok2013, birkby2013, piskorz2016}. Comparison has also shown that some methods are better at correcting H$_2$O than O$_2$ \citep{ulmer-moll2019}.

In this study we investigate how different telluric correction methods may affect the measurements of chemical species in exoplanet atmospheres at optical wavelengths. We see inconsistencies in measured results depending on the method employed and the nightly weather variation, so it is important to quantify which techniques are best for analysing high-resolution exoplanet spectra. We choose to compare two popular methods: one which derives the telluric transmission solely from the data, and one which uses a model of molecular absorption in Earth's atmosphere. Both of these methods account for the variation of telluric line strength over the multiple-hour observing window. This gives an advantage over modelling a telluric standard star, where the resultant spectrum is insensitive to atmospheric variation during the transit and requires extra time to observe. We focus specifically on the quality of telluric corrections in the region around the sodium D-lines, however the analysis is applicable to the full range of the spectrograph. Other methods discussed above may be more suited for Doppler-resolved molecular detections (particularly in the near-infrared) but were beyond the scope of this work.

We focus on HD~189733~b -- a tidally locked hot Jupiter orbiting a bright K1V host star with magnitude V = 7.67. There have already been a number of atmospheric detections for this planet: 
Na \citep{redfield2008, jensen2011, huitson2012, wyttenbach2015, khalafinejad2017}, 
H$_2$O \citep{birkby2013, mccullough2014, brogi2016, brogi2018, cabot2019}, 
H \citep{jensen2012, lecavelierderetangs2010, bourrier2013}, 
CO \citep{dekok2013, brogi2016, cabot2019}, 
He \citep{salz2018}, and 
HCN \citep{cabot2019}. 
Additionally, there has been evidence of Rayleigh scattering and high-altitude hazes \citep{pont2008, lecavelierdesetangs2008, digloria2015}, and measurement of atmospheric wind speeds \citep{louden2015}.

In section~\ref{sec:observations} we give an overview of the high-resolution HARPS spectra used for our analysis. We discuss the data reduction and two methods for removing telluric contamination in section~\ref{sec:reduction}, and the calculation of the transmission spectrum which is common for both methods in section~\ref{sec:method_transmission}. In section~\ref{sec:results} we evaluate the differences in the telluric reduction processes by measuring the absorption parameters of the Na doublet using Gaussian fits and a binned passband analysis. We also measure the net atmospheric wind speed from the blueshift of the line profiles and discuss further atmospheric properties. The results are briefly summarised in section~\ref{sec:conclusion} together with a conclusion about the most advantageous method.

\section{Observations}
\label{sec:observations}

\begin{table}
    \centering
    \begin{tabular*}{\columnwidth}{c@{\extracolsep{\fill}}ccc}
    \hline
    Night & Date & \# In/Out Spectra & Program ID \\
    \hline
    0 & 2006-07-29 & 6/6 & 072.C-0488(E) \\
    1 & 2006-09-07 & 10/10 & 072.C-0488(E) \\
    2 & 2007-07-19 & 18/21 & 079.C-0828(A) \\
    3 & 2007-08-28 & 20/20 & 079.C-0127(A) \\
    \hline
    \end{tabular*}
    \caption{HARPS observing log of HD~189733. Nights 1--3 are numbered to remain consistent with \citet{wyttenbach2015}. The number of in and out-of-transit exposures are calculated using the mid-exposure times and the ephemeris in Table~\ref{tab:parameters}.}
    \label{tab:obs_log}
\end{table}

\begin{figure}
    \centering
    \includegraphics[width=\columnwidth]{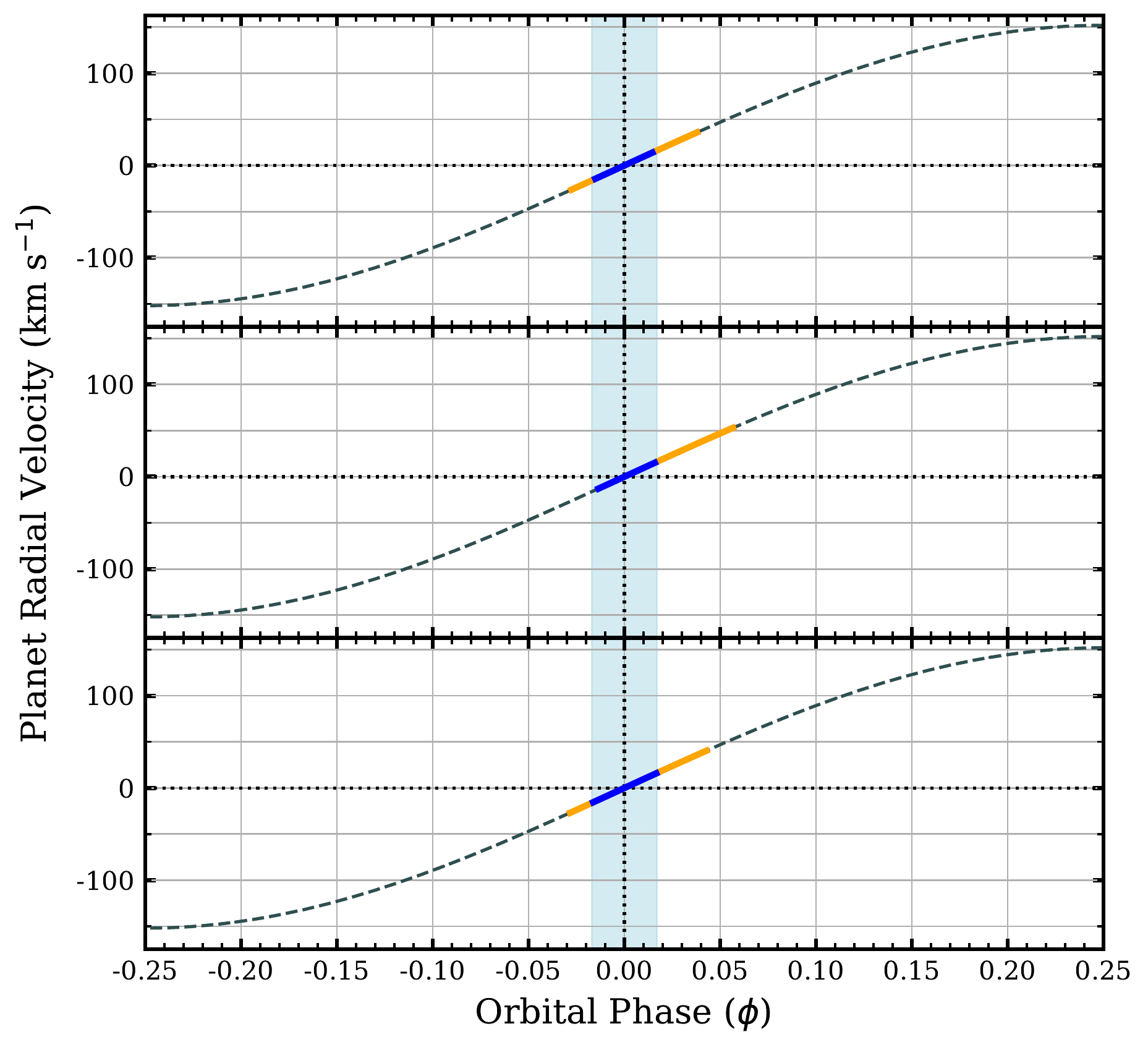}
    \caption[Planetary RV Curve]{Planetary radial velocity curves for observations on night 1 (top), 2 (middle), and 3 (bottom). The dashed line shows the modelled radial velocity curve, and the phase coverage of in and out-of-transit observations are shown with blue and orange lines respectively. The transit window (time between first and fourth contact) is shaded.}
    \label{fig:phases}
\end{figure}

In this paper we use observations of high-resolution transmission spectra of the hot Jupiter HD~189733~b as a case study. Data were obtained from programs 072.C-0488(E), 079.C-0828(A) (PI: Mayor) and 079.C-0127(A) (PI: Lecavelier des Etangs) using the High Accuracy Radial velocity Planet Searcher (HARPS) echelle spectrograph on the ESO 3.6~m telescope in La Silla, Chile, and made available through the ESO archive. With a resolving power of $R =$~115,000, HARPS records 72 orders of the echelle spectrum over a range of 380--690~nm. The detector contains a mosaic of two 4k~x~4k pixel CCDs; one spectral order from 530--533~nm is lost due to a gap between the CCDs. Two fibres are used when observing -- fibre A for the target and fibre B for recording simultaneous reference spectra for calibrations. Each fibre has a diameter of 70~$\micron$ which gives a 1~arcsec aperture on the sky \citep{mayor2003}. All observations are reduced by the HARPS Data Reduction Software (DRS) v3.5. The pipeline performs calibrations using the dark, bias and flat-field frames taken at the beginning of the night, corrects for the blaze, and uses Thorium-Argon (Th-Ar) reference exposures for wavelength calibration. Finally, each spectrum is remapped onto a uniform grid at a resolution of 0.01~{\AA} in the solar system barycentric rest frame.

High-resolution spectra of HD~189733 were obtained with HARPS during four primary transits and are available through the ESO archive. The observing log in Table~\ref{tab:obs_log} shows the dates and program IDs of the observations. The number of in-transit spectra was calculated using the mid-exposure time and ephemeris of \citet{agol2010}. A combination of these datasets was first used to measure the RM effect \citep{triaud2009}. They have since been analysed by several other authors to study other atmospheric properties, leading to important results such as:
a detection of sodium in the planetary atmosphere \citep{wyttenbach2015}, 
atmospheric winds \citep{louden2015, seidel2020a}, 
a temperature gradient of ${\sim}0.4$~K~km$^{-1}$\citep{heng2015}, 
a measurement of the Rayleigh scattering slope \citep{digloria2015}, 
and an attempt to constrain the presence of atmospheric water vapour \citep{allart2017}.
As discussed by many of these authors, on 29 July 2006 there are only 12 observations in total, and there are none during the second half of the transit because of bad weather conditions. We were unable to recover a clear planetary transmission spectrum due to the lack of data before and during transit, and therefore chose to ignore this night.

Parameters of the HD~189733 system which were used throughout this study are shown in Table~\ref{tab:parameters}. The known stellar radial velocity semi-amplitude was used to model the stellar and planetary radial velocity curves. Spectra with mid-exposure times between the first and fourth contacts of the eclipse are defined as in-transit. As seen in Figure~\ref{fig:phases}, the observations on nights 1 and 3 fully cover the period shortly before, during, and after the transit. The orange and blue lines indicate out of and in-transit observations respectively. The complete transit duration is shown by the shaded region.

\section{Data Reduction}
\label{sec:reduction}

Preliminary corrections need to be made to the spectra before telluric contamination can be considered. We extract the data from the one-dimensional HARPS (s1d) files. The full spectral wavelength range is 3781--6912~{\AA}, however we limit this to 4000--6800~{\AA} to reduce systematics at the edges of the full spectrum where strong telluric contamination or low throughput is evident. A small number of pixels in each spectrum contain substantially higher flux values than the median, which are unlikely to be from the observed astrophysical source. To correct for these, a mask is created for each spectrum by applying a median filter with a width of 9 pixels and subtracting this from the original spectrum. Any pixels in the mask which are greater than 10 standard deviations from the median are flagged, and the corresponding fluxes in the original spectra are corrected to the median of the 10 surrounding pixels. Each spectrum is then normalised by dividing by its median. The errors in the flux values are assumed to be dominated by photon noise and we propagate the Poisson uncertainties throughout the analysis.

All ground-based observations in this wavelength range are polluted with telluric lines predominantly due to H${_2}$O and O$_2$. These lines vary in strength with observing conditions such as airmass and precipitable water vapour. Incorrect removal of tellurics will create false signals in the resultant planetary transmission spectrum. We now consider the removal of telluric contamination using two methods.

\subsection{Correcting tellurics with airmass}
\label{sec:reduction_airmass}

We first employed a method similar to that described by \citet{vidal-madjar2010}, \citet{astudillo2013}, and \citet{wyttenbach2015} -- by assuming that telluric lines have a linear variation with airmass. The airmass at the start and end of each observation is measured on-site and stored in the headers of the HARPS files -- we assigned an average airmass to each observation using these two measurements. Choosing only the starting or ending value caused a slight difference in the telluric correction. For each wavelength step in the spectrum, a linear fit between the natural logarithm of the measured fluxes and their corresponding airmass was made, taking the form
\begin{equation}
    \text{ln}(F(\lambda)) = (Nk_\lambda)s + c ~,
    \label{eqn:linearfit}
\end{equation}
where $F(\lambda)$ is the flux at wavelength $\lambda$, $Nk_\lambda$ is the zenithal optical depth, and $s$ is the airmass. The telluric lines vary with airmass but the exoplanet absorption lines do not. Therefore a telluric reference spectrum, $T(\lambda)$, can be built using the value of the gradient in equation~\ref{eqn:linearfit} \citep{wyttenbach2015}
\begin{equation}
    T(\lambda) = \text{e}^{Nk_\lambda} ~.
    \label{eqn:telluric}
\end{equation}
It is important to only use out-of-transit spectra to build the telluric reference spectrum; using in-transit spectra would also correct for absorption due to the planet's atmosphere. Therefore, it is ideal to obtain a number of spectra shortly before and shortly after the transit to ensure a large enough out-of-transit baseline. However, observations on night 2 began when the planet was already in transit and there are not enough out-of-transit observations to produce a good-quality telluric reference spectrum. We were unable to recover a clear transmission spectrum using the limited out-of-transit data, so chose to include the in-transit sample as well. We note that some of the planetary signals could be overcorrected and muted as a result.

\begin{figure}
    \centering
    \includegraphics[width=\columnwidth]{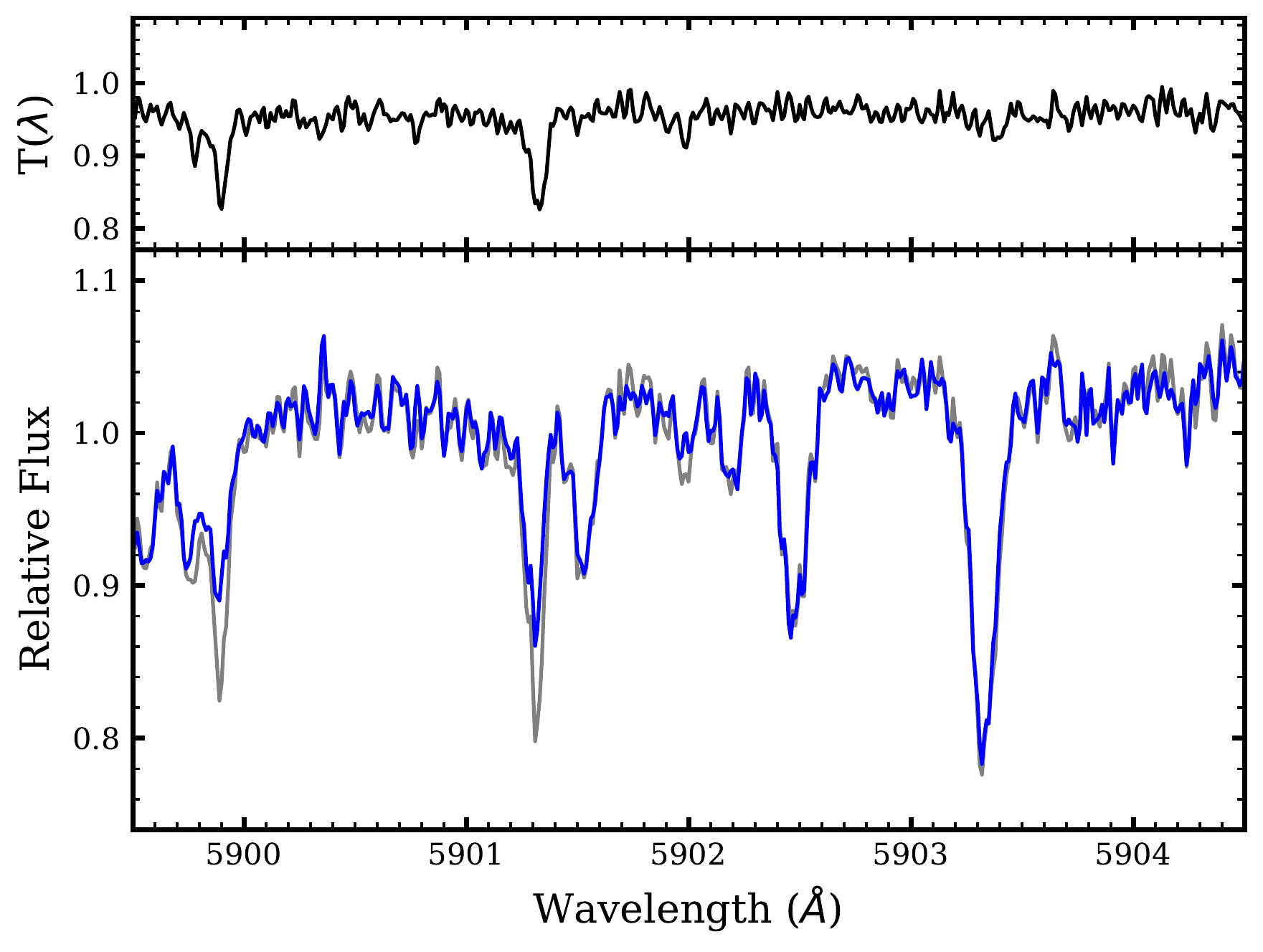}
    \caption{The effect of re-scaling the strength of telluric lines using the empirical telluric reference spectrum. \textit{Top panel:} derived telluric spectrum with prominent features around 5899.9 and 5901.3~{\AA}. \textit{Bottom panel:} uncorrected (grey) and re-scaled data (blue). The telluric lines in each spectrum are reduced so that they appear as though observations took place with a constant airmass. Stellar lines remain unchanged.}
    \label{fig:airmass_correction}
\end{figure}

Using equation~\ref{eqn:telluric}, the telluric contamination in the observed spectra can be adjusted so that they appear to have been observed at a constant reference airmass, $s_{\text{ref}}$. The adjusted spectra, $F_{\text{ref}}$, were calculated for all observations using
\begin{equation}
    F_{\text{ref}}(\lambda) = \frac{F_{\text{obs}}(\lambda)}{T(\lambda)^{(s - s_{\text{ref}})}} ~,
    \label{eqn:Fref}
\end{equation}
where $F_{\text{obs}}(\lambda)$ is the observed spectrum and $s$ is the airmass at the time of the observation \citep{wyttenbach2015}. The reference airmass was taken to be the mean airmass of the in-transit sample only. An example of the adjustment to telluric lines is shown in Figure~\ref{fig:airmass_correction}. The top panel shows the calculated telluric reference spectrum $T(\lambda)$, and the bottom panel shows the result of the correction with the original data in grey and the corrected data in blue. It is noted that the telluric lines are only adjusted to a level of constant airmass but are not removed completely. Following this correction, the transmission spectrum was computed as described in section~\ref{sec:method_transmission}.

Residuals of telluric lines may occasionally be seen in the transmission spectrum despite performing the correction derived from airmass. These lines may be confused for planetary atmospheric absorption, but are actually caused by changes in Earth's atmospheric water content in the air above the telescope throughout the night. Similarly to \citet{wyttenbach2015}, we make a second correction to account for this. For the data on night~3, a linear fit between the transmission spectrum and $T(\lambda)$ was made. The transmission spectrum was divided by this fit which effectively reduces parts of the spectrum where telluric residuals are correlated with $T(\lambda)$. We split up the full spectrum into smaller wavelength ranges and performed the correction on each section to prevent variations in unrelated parts of the spectrum from affecting each other. Only one iteration of the correction was needed to reduce the residuals to the continuum level. This is further discussed in section~\ref{sec:results_telluric_removal}.

\subsection{Correcting tellurics with \texttt{molecfit}}
\label{sec:reduction_molecfit}

An ideal method would reduce telluric contamination down to the noise level without changing any of the existing data outside of these regions. The empirical method discussed in the previous section struggles to do this, and gets worse with low signal-to-noise.

Here we consider telluric removal by modelling out Earth's atmospheric contribution. We use \texttt{molecfit} v1.2.0 -- an ESO tool specifically designed to correct for telluric contamination in ground-based spectra \citep{smette2015, kausch2015}. \texttt{Molecfit} fits a line-by-line radiative transfer model (LBLRTM) of telluric absorption to the observed spectra, producing a unique model of atmospheric absorption for each observation. This was first performed on HARPS spectra by \citet{allart2017}, and has since been used in other HARPS studies \citep{cauley2017, seidel2019, cabot2020}, as well as with other high-resolution spectrographs \citep{cauley2017, cauley2019, allart2018, allart2019, nortmann2018, salz2018, casasayas-barris2019, casasayas-barris2020, hoeijmakers2019, alonso-floriano2019, kirk2020, chen2020}.

\begin{figure*}
    \centering
    \includegraphics[width=\textwidth]{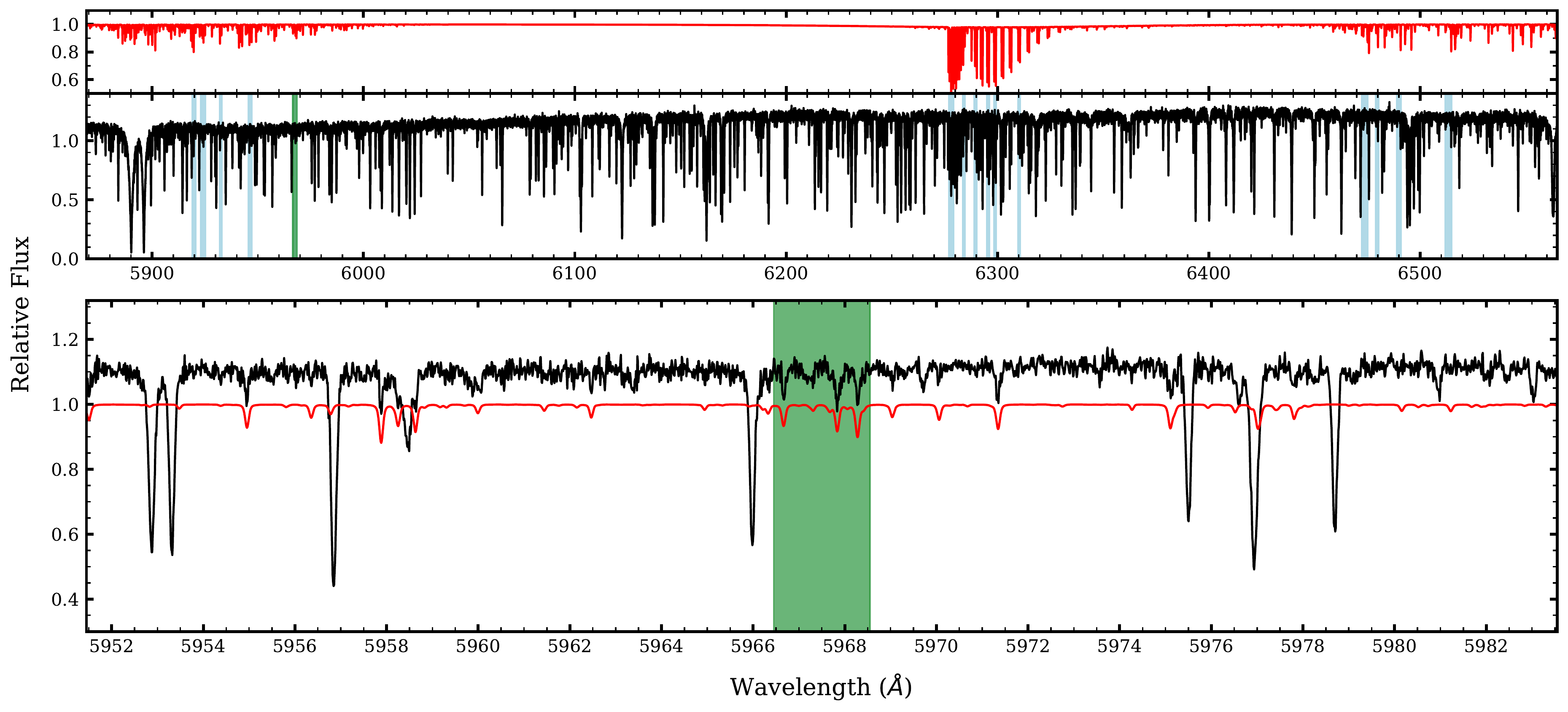}
    \caption{Example of parts of the spectra which are contaminated with telluric lines. The 15 shaded light blue/green regions are passed to \texttt{molecfit}. \textit{Top panel:} modelled telluric spectrum (red) with H$_2$O absorption in bands around ${\sim}5900$~{\AA} and ${\sim}6500$~{\AA}, and O$_2$ absorption around ${\sim}6300$~{\AA}. \textit{Middle panel:} data (black) and the chosen regions (shaded) with widths <2~{\AA}. \textit{Bottom panel:} expanded view of the green shaded region from the middle panel: by comparison with the \texttt{molecfit} model (red), we make sure that each region only contains telluric lines with no other stellar features.}
    \label{fig:molecfit_regions}
\end{figure*}

HARPS spectra are given in the Solar System barycentric rest frame. We use the Barycentric Earth Radial Velocity (BERV) values to shift each spectrum into the rest frame of the telescope to ensure correct telluric modelling with \texttt{molecfit}. We then inspect areas with heavy H$_2$O and O$_2$ absorption around 5950, 6300, and 6475~{\AA}, and select 15 small regions (no larger than 2~{\AA}) which contain only telluric lines and flat continuum (Figure~\ref{fig:molecfit_regions}). It is important that no stellar lines are included as they will affect the fit to the atmospheric model. The chosen regions remain constant for the duration of one night of observing, however they must be re-selected for different nights since the location of telluric lines changes with respect to the stellar lines.

\begin{figure*}
    \centering
    \includegraphics[width=\textwidth]{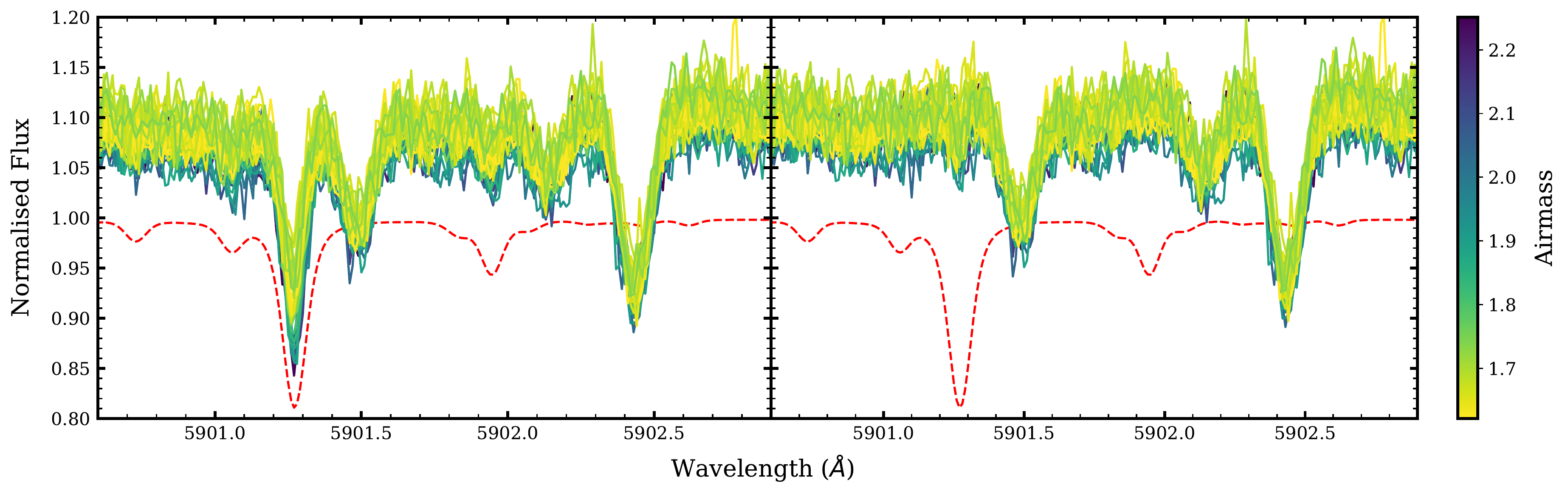}
    \caption{Result of telluric corrections with \texttt{molecfit}. The spectra for night 3 are overlaid and coloured according to their respective airmass, and the red dashed line shows the telluric model for the first observation. \textit{Left panel:} uncorrected data with clear telluric contamination around 5901.3~{\AA}. \textit{Right panel:} corrected data following division of the model. The telluric lines have been reduced to the noise level.}
    \label{fig:telluric_molecfit}
\end{figure*}

We provide \texttt{molecfit} with parameters similar to those discussed by \citet{allart2017}, together with the date, location, and atmospheric conditions (e.g. humidity, pressure, temperature) stored within the HARPS output. The atmospheric model is fitted to the selected regions. With reference to the output parameters, we then run the \texttt{calctrans} tool which fits an atmospheric model to the entire spectrum, resulting in a unique telluric profile for each spectrum with the same resolution. Finally, the spectra are divided by their corresponding model to reduce all telluric lines down to the continuum noise level, as shown in Figure~\ref{fig:telluric_molecfit}.

\section{Transmission spectra}
\label{sec:method_transmission}

Here we discuss our calculation of the transmission spectra from observations. 
This work addresses several common effects that can imprint false signatures onto the planetary transmission spectrum at high-resolution. These effects include telluric contamination, velocity contributions from stellar, systemic and planetary sources, as well as spurious signals due to the Rossiter-McLaughlin (RM) effect and Centre-to-Limb Variation (CLV). 

\subsection{Velocity corrections}
Following removal of telluric contamination (sections~\ref{sec:reduction_airmass} and \ref{sec:reduction_molecfit}), we extract the transmission spectrum while making corrections for the stellar reflex motion, systemic velocity, and planetary radial velocity using a similar technique to \citet{wyttenbach2015}. Individual spectra are identified as $f(\lambda,t_{\text{in}})$ for in-transit exposures and $f(\lambda,t_{\text{out}})$ for out-of-transit exposures by modelling the orbit using the parameters listed in Table~\ref{tab:parameters}. As discussed in section~\ref{sec:observations}, it is important that the observations cover the period shortly before, during, and shortly after the transit to create a good out-of-transit baseline.

\begin{figure}
    \centering
    \includegraphics[width=\columnwidth]{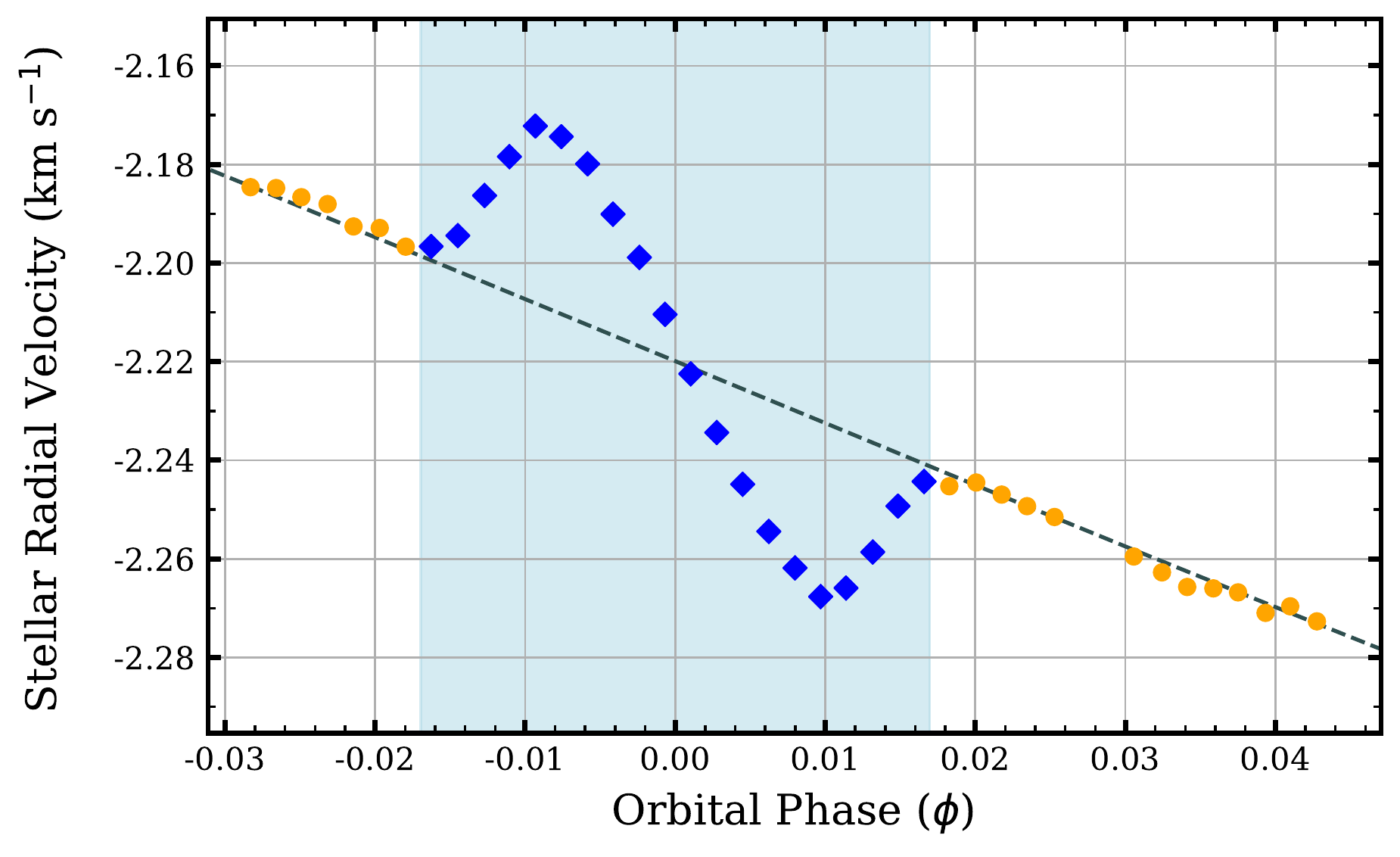}
    \caption{Stellar radial velocity measurements for night 3, as measured by HARPS. In-transit observations are shown with blue diamonds, and out-of-transit with orange circles. Deviation from the modelled radial velocity curve (dashed line) during the shaded in-transit window is due to the Rossiter-McLaughlin effect.}
    \label{fig:stellar_rv}
\end{figure}

The HARPS-measured stellar radial velocities at the time of each exposure are extracted from the Cross Correlation Function (CCF) files. An example of these measurements for observations on night~3 is shown in Figure~\ref{fig:stellar_rv}. The dashed line is a fit for both the stellar reflex motion and the systemic velocity, taking the form
\begin{equation}
    v_{\ast} = K_{\ast}\sin{(2\pi\phi)} +v_{\text{sys}} ~,
    \label{eqn:stellar_RV}
\end{equation}
where $K_{\ast}$ is the stellar radial velocity semi-amplitude, $\phi$ is the phase and $v_{\text{sys}}$ is the systemic velocity. For the in-transit spectra, there is a deviation from this fit due to the RM effect \citep{triaud2009, digloria2015}. Each spectrum is Doppler shifted to correct for its corresponding measured stellar radial velocity and linearly interpolated back to the uniform 0.01~{\AA} grid. This corrects for the apparent radial velocity change due to the RM effect, however the overall shape of the transmission line profiles will still be affected and could lead to spurious signals -- additional corrections are discussed in section~\ref{sec:method_RM}. We simultaneously calculate the planetary radial velocity for all observed phases using 
\begin{equation}
    v_{\text{p}} = K_{\text{p}}\sin{(2\pi\phi)} ~,
    \label{eqn:planet_RV}
\end{equation}
where $K_{\text{p}} = -K_{\ast}(M_{\ast}/M_{\text{p}})$.

The maximum change in measured stellar radial velocity throughout one night of observing is ${\sim}90$~ms$^{-1}$, corresponding to a ${\sim}0.0018$~{\AA} wavelength shift. The planetary radial velocity is much larger in comparison, varying between $\pm{16}$~kms$^{-1}$ during the transit. As a result, signals in the planetary transmission will be shifted up to ${\sim}0.32$~{\AA} (32 resolution elements) from the rest wavelength. It is therefore important to correct for this in order to recover the full transmission spectrum.

We propagate the Poisson uncertainties of the raw data and use the inverse variance as weights when combining the spectra. This reduces the contribution of spectra with low signal-to-noise which may have been affected by poor weather conditions or other time-dependent systematics. We co-add all out-of-transit exposures to create a master out-of-transit spectrum, denoted $\overline{f}_{\text{out}}(\lambda)$. In-transit spectra contain signals from both the star and planet, whereas the out-of-transit spectra contain just the stellar signal. Classically, the transmission spectrum can be calculated by dividing a master in-transit spectrum, $\overline{f}_{\text{in}}(\lambda)$, by $\overline{f}_{\text{out}}(\lambda)$, however this does not account for the planetary radial velocity shift. We therefore compute individual transmission spectra as
\begin{equation}
    \Re(\lambda,t_{\text{in}}) = \frac{f(\lambda,t_{\text{in}})}{\overline{f}_{\text{out}}(\lambda)} ~.
    \label{eqn:transmission_individual}
\end{equation}
Each spectrum is then continuum normalised using a third-order polynomial to remove effects from instrumental systematics or weather variations. A Doppler shift using $v_{\text{p}}$ is performed to move each spectrum into the planet's rest frame, and the final combined transmission spectrum is thus computed:
\begin{equation}
    {\Re'}(\lambda) = \sum_{t_{\text{in}}}\Re(\lambda,t_{\text{in}})|_{v_{\text{p}}(t_{\text{in}})} - 1 = \frac{F_{\lambda, \text{in}}}{F_{\lambda, \text{out}}} -1 ~,
    \label{eqn:transmission_combined}
\end{equation}
where $F_{\lambda, \text{in}}$ is the Doppler-corrected and time-averaged in-transit spectrum.
We remove any remaining broadband continuum variations with a median filter of width 1501 pixels (15.01~{\AA}).

\subsection{Correcting CLV and RM effects}
\label{sec:method_RM}

CLV and RM effects can change the shape of lines in the stellar spectrum during the transit. This induces spurious signals into the transmission spectrum which must be corrected for \citep{queloz2000, triaud2018}. We model the extent of these effects in the resultant transmission spectra in a similar manner as \citet{casasayas-barris2019}. We generate synthetic spectra of HD~189733 using the software {\texttt{Spectroscopy Made Easy (SME)}} \citep{Valenti1996}. The synthesis routine takes as input: (1) physical stellar parameters, for which we used those listed in Table~\ref{tab:parameters} (we do not provide a stellar rotation at this point); (2) a list of spectral line positions and strengths, for which we use the VALD3 database \citep{Ryabchikova2015}; and (3) a list of $\mu$-angles \citep{Mandel2002}; we choose 21 angles spanning from the limb to the centre of the stellar disk. For each night of observation, we simulate the transit of the planet on an $80\times80$ pixel grid centred on the host star. Each pixel overlapping the stellar disc is assigned a synthetic spectrum, interpolated at the appropriate $\mu$-angle from the set of calculated {\texttt{SME}} spectra. The spectrum is also Doppler shifted according to the local radial velocity of the star, per parameters in Table~\ref{tab:parameters}. The integrated flux over all pixels provides our simulated out-of-transit spectrum. At each phase of the transit, corresponding to exposure midtimes, we geometrically determine the projected position of the planet on the stellar disc \citep{cegla2016}, and integrate the stellar flux over all pixels not occulted by the planet. The result is our simulated in-transit spectra. The jointly modelled CLV and RM effect on the observed transmission spectra is obtained by dividing each simulated in-transit spectrum by the simulated out-of-transit spectrum. We do not account for potential non-LTE effects in the cores of the Na doublet, nor variable planetary radii accounting for absorption by the atmosphere.

\begin{figure}
    \centering
    \includegraphics[width=\columnwidth]{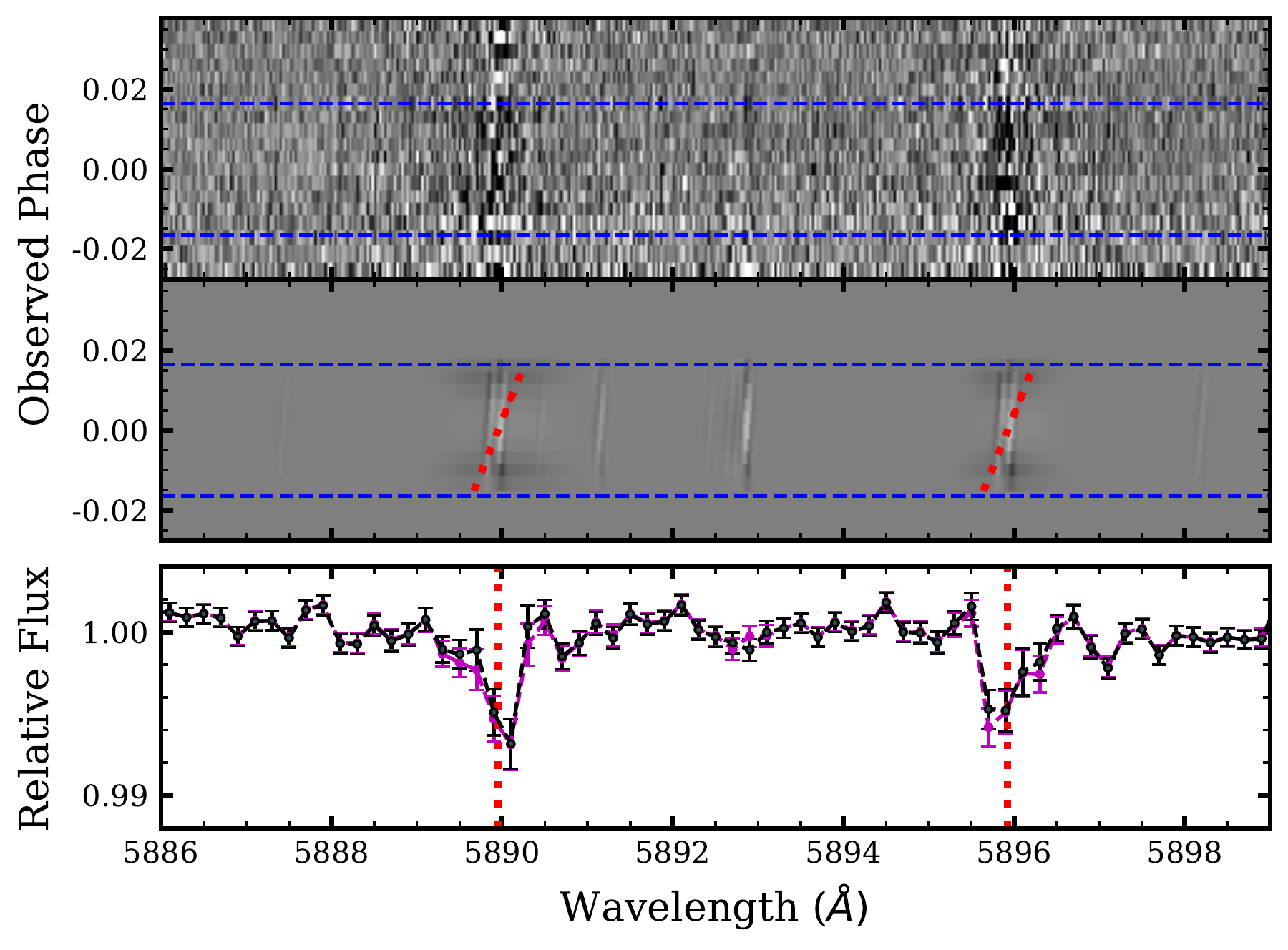}
    \caption{Comparison of the transmission spectrum and the modelled CLV and RM effects around the Na doublet for night 1. The colour scale in panels 1 and 2 represents relative flux from 0.98 (black) to 1.02 (white), and the blue dashed lines indicate the transit window. \textit{Top panel:} planetary transmission spectrum for each observed phase. \textit{Second panel:} modelled transmission spectrum assuming no atmosphere, showing artefacts of the CLV and RM effects. \textit{Third panel:} binned (20$\times$) CLV/RM uncorrected (magenta) and corrected (black) transmission spectrum in the planetary rest frame. The red dotted lines in panels 2 and 3 give the position of the sodium doublet in the planetary rest frame.}
    \label{fig:rm_correction}
\end{figure}

As shown in the middle panel of Figure~\ref{fig:rm_correction}, the modelled transmission spectrum contains features which arise strictly from CLV and RM effects. However, these features have amplitudes $\lesssim0.001$ relative flux. This is small in comparison to the planetary absorption signal (discussed in the following section), and approximately at the noise level of the data. Indeed, following the Na doublet detection by \citet{wyttenbach2015}, further studies which account for the RM correction \citep{louden2015, casasayas-barris2017} detect the lines at similar strengths. The consistency may be explained by the fact that HD~189733 is a slow rotator, and the occulted, local radial velocities differ significantly from the planet's Doppler motion traced by the atmospheric absorption. Still, these studies find a several km s$^{-1}$ smaller net blueshift in the Na doublet which underscores the potential impact of CLV and RM residuals, as well as the RM velocimetric effect. We correct for small changes in the shape of the line profiles by dividing by the model, and produce a final, observed transmission spectrum by co-adding the overall nightly CLV and RM corrected spectra.

\section{Results and Discussion}
\label{sec:results}

In this section we show a comparison of the calculated HD~189733~b transmission spectra using the two different telluric corrections. We confirm and discuss the detection of Na \textsc{i} in the planetary atmosphere, assess the shape of spectral features and the amount of absorption, and give context for further inferred atmospheric properties. Since we are focusing on the Na doublet, we restrict the wavelength range of the transmission spectra to 5870--5916~{\AA}.

\subsection{Telluric removal methods}
\label{sec:results_telluric_removal}

\begin{figure}
    \centering
    \includegraphics[width=\columnwidth]{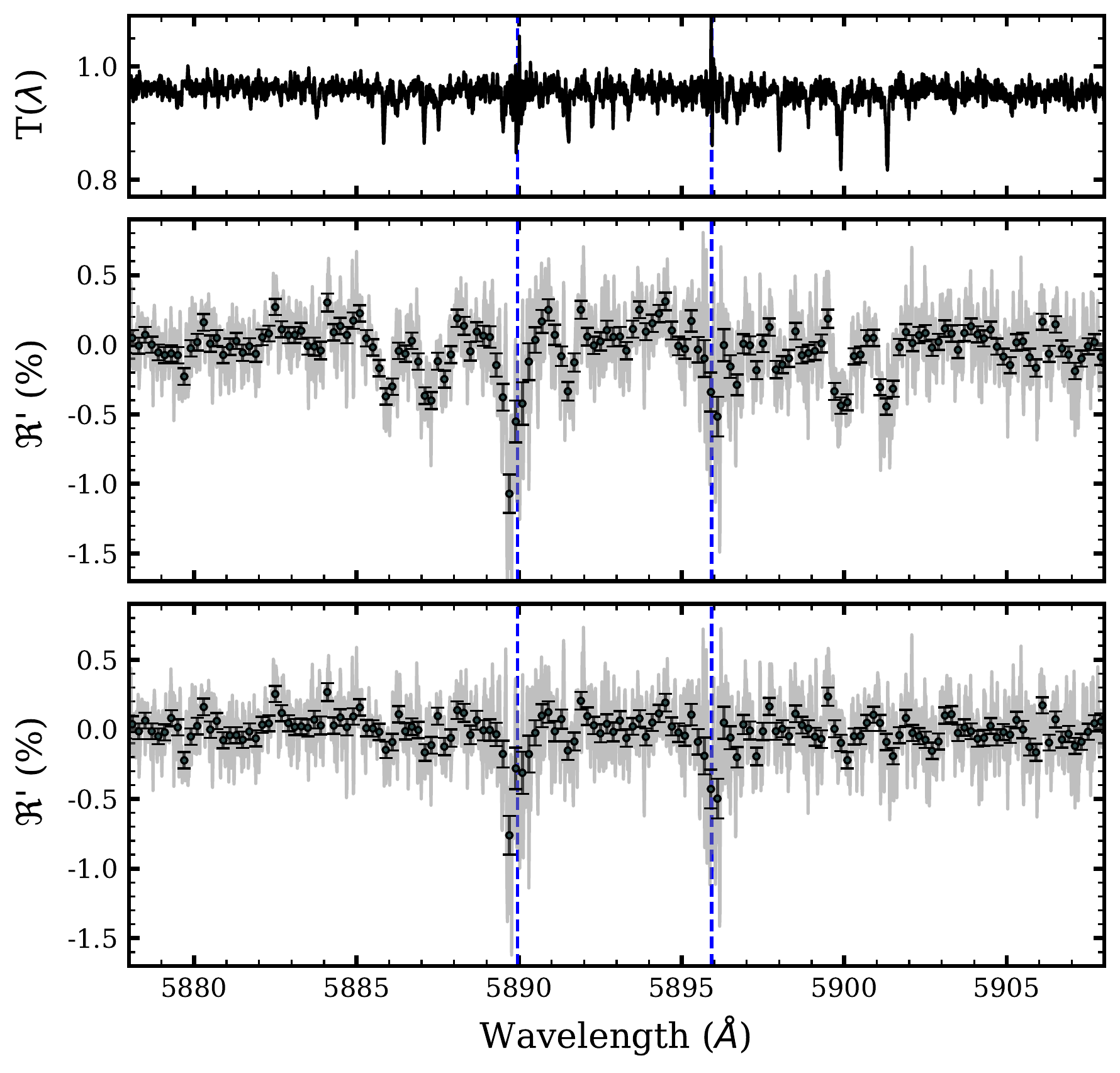}
    \caption{Effect of telluric corrections on the transmission spectrum for night 3. \textit{Top panel:} empirical telluric reference spectrum. \textit{Middle panel:} transmission spectrum after airmass telluric corrections. \textit{Bottom panel:} transmission spectrum and after applying a second correction for water column variation. The full-resolution spectra are shown in grey, and binned 20$\times$ in black. By comparing the middle panel with the derived telluric reference spectrum (top panel), residual telluric lines can be seen in regions around 5886.0, 5887.5, 5891.5, 5900.0, and 5901.5~{\AA}. These are removed following the extra correction.}
    \label{fig:transmission_water}
\end{figure}

For a clearer visualisation of the spectral features, the transmission spectra for all nights were binned to a 0.2~{\AA} resolution (20 points per bin). We define absorption features as parts of the spectrum which are negative in value. Figure~\ref{fig:transmission_water} shows the transmission spectrum for night 3 using the airmass correction method. On this night, a secondary correction was performed to reduce the effects of water column variation. The middle panel shows the transmission spectrum without the extra correction -- by comparing this with the empirical telluric reference spectrum (top panel), it is clear that there are still residual telluric lines present around 5886.0, 5887.5, 5891.5, 5900.0, and 5901.5~{\AA}. Since these are of comparable strengths to the signals we expect in the planetary transmission spectrum, the validity of any detections is significantly reduced. The bottom panel shows the result after applying the second correction, and the residual tellurics have been reduced without affecting other parts of the spectrum. However, the sodium lines could be slightly overcorrected due to increased noise in the telluric reference spectrum in these regions. This correction was not necessary for nights 1 and 2.

\begingroup
\renewcommand{\arraystretch}{1.2} 
\begin{table}
    \centering
    \begin{tabular*}{\columnwidth}{c@{\extracolsep{\fill}}ccc}
    \hline
    Line                    & $\lambda_0$ ({\AA})   & $\mathcal{D}$ (\%)    & FWHM ({\AA})      \\
    \hline
    D2\textsubscript{a}     & $5889.88 \pm 0.04$    & $-0.55 \pm 0.07$      & $0.46 \pm 0.05$   \\
    D1\textsubscript{a}     & $5895.94 \pm 0.04$    & $-0.45 \pm 0.07$      & $0.40 \pm 0.06$   \\
    \hline
    D2\textsubscript{m}     & $5889.87 \pm 0.03$    & $-0.64 \pm 0.07$      & $0.46 \pm 0.04$   \\
    D1\textsubscript{m}     & $5895.93 \pm 0.04$    & $-0.53 \pm 0.07$      & $0.41 \pm 0.05$   \\
    \hline
    \end{tabular*}
    \caption{Measured properties of the Gaussian fits to the sodium D-lines in the combined transmission spectrum of HD~189733~b. Lines are labelled with subscripts 'a' and 'm' to denote which telluric removal method (airmass or \texttt{molecfit}) was used.}
    \label{tab:gauss_fit}
\end{table}
\endgroup

To assess the quality of the telluric reduction methods, we fit Gaussian profiles to the lines in the sodium doublet which provides a good approximation of the line cores \citep{gandhi2017}. We used the \texttt{LevMarLSQFitter} module from \texttt{astropy}, which performs a least-squares fit using a Lavenberg-Marquardt algorithm and accounts for errors in the flux which have been propagated from the photon noise on the raw spectra. We define three properties of the Gaussian fit: the amplitude or line contrast, $\mathcal{D}$; centroid $\lambda_0$; and the full width at half maximum (FWHM). Figure~\ref{fig:transmission_n3} shows the night 3 transmission spectrum around the Na doublet for the airmass (left) and \texttt{molecfit} (right) telluric removal methods. By comparison, we see that corrections using airmass gives a transmission spectrum with more variation, particularly in regions where residual telluric lines were removed (see Figure~\ref{fig:transmission_water}). The Gaussian fits to the full-resolution data around the Na lines have reduced chi-square values of $\chi^2_\nu = 0.54$ for the airmass correction method and $\chi^2_\nu = 0.50$ for the \texttt{molecfit} method.

For each method, we co-added the nightly spectra to ensure that the signal-to-noise ratio is large enough for a well-characterised atmospheric sodium detection. Weaker signals would also be more prominent. The final CLV and RM corrected transmission spectra are shown in Figure~\ref{fig:transmission_combined}. The Gaussian fits have $\chi^2_\nu = 0.58$ (airmass) and $\chi^2_\nu = 0.57$ (\texttt{molecfit}). Given the comparable fits with both approaches, no conclusion can be drawn about which method is better based on this statistic alone. The measured parameters of the Gaussian fits to the sodium D2 and D1 lines are shown in Table~\ref{tab:gauss_fit}. Following reduction using the airmass telluric correction method, we measured line contrasts of $-0.55 \pm 0.07~\%$ for the D2 line and $-0.45 \pm 0.07~\%$ for the D1 line. This results in an average depth of $-0.50 \pm 0.05~\%$. For the \texttt{molecfit} method, the line contrasts were $-0.64 \pm 0.07~\%$ (D2) and $-0.53 \pm 0.07~\%$ (D1), averaging to $-0.59 \pm 0.05~\%$. Although the average line contrasts for both methods are not within 1$\sigma$ of each other, the error ranges overlap. From this we note that telluric corrections using the \texttt{molecfit} method resulted in a transmission spectrum with deeper Na features than those for the airmass method. We find a similar difference when measuring binned absorption depths, and further context is provided in section~\ref{sec:results_depths}. The measured line contrasts and differences between the D2 and D1 lines are comparable to the results of \citet{wyttenbach2015} (D2:~$-0.64~\pm~0.07~\%$, D1:~$-0.40~\pm~0.07~\%$) and \citet{casasayas-barris2017} (D2:~$-0.72 \pm 0.05~\%$, D1:~$-0.51 \pm 0.05~\%$).

\begingroup
\begin{figure*}
    \centering
    \includegraphics[width=\textwidth]{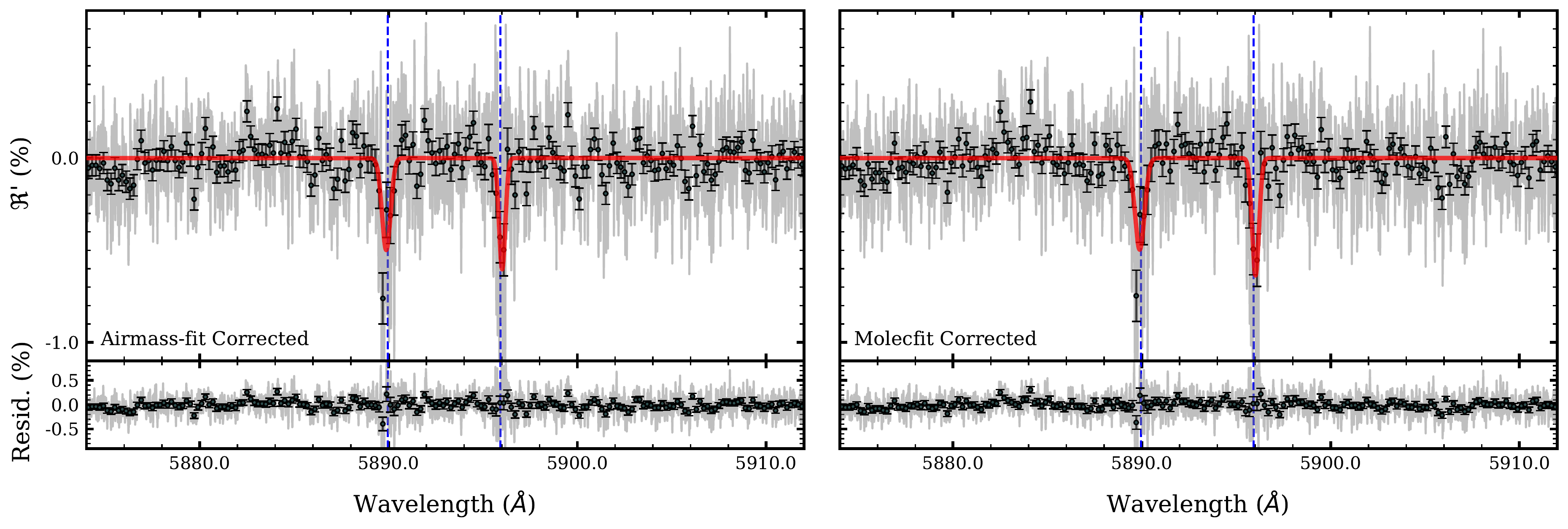}
    \caption{Na \textsc{i} doublet absorption in the atmosphere of HD~189733~b. The top row shows the transmission spectrum for night 3 (in the planetary rest frame) using the airmass (left) and \texttt{molecfit} (right) telluric correction methods. Full-resolution data are shown in grey, and binned 20$\times$ in black. The bottom row shows the residuals to the best-fitting Gaussian profiles (red) in the same resolution. The expected rest frame positions of the D-lines are indicated with the blue dashed lines.}
    \label{fig:transmission_n3}
\end{figure*}

\begin{figure*}
    \centering
    \includegraphics[width=\textwidth]{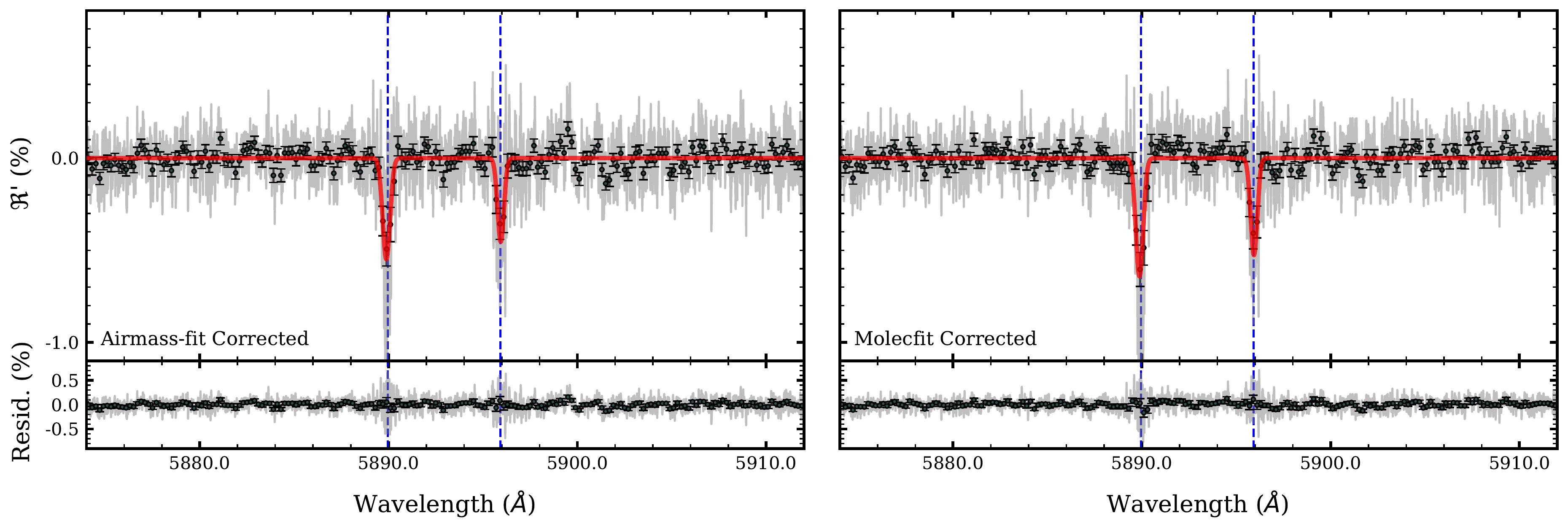}
    \caption{Same as Figure~\ref{fig:transmission_n3} but for the combined data across all three nights. Stacking multiple nights of data in the rest frame of the planet increases the signal-to-noise ratio, as is evident by visual comparison with the transmission spectrum for night 3.}
    \label{fig:transmission_combined}
\end{figure*}

\renewcommand{\arraystretch}{1.2} 
\begin{table*}
    \centering
    \begin{tabular*}{\textwidth}{c@{\extracolsep{\fill}}cccccc}
    \hline
    Night       & AD$^{0.375}_{\text{a}}$   & AD$^{0.75}_{\text{a}}$    & AD$^{1.5}_{\text{a}}$     & AD$^{3}_{\text{a}}$       & AD$^{6}_{\text{a}}$       & AD$^{12}_{\text{a}}$  \\
    \hline
    1           & $-0.556 \pm 0.102$        & $-0.632 \pm 0.072$        & $-0.428 \pm 0.049$        & $-0.260 \pm 0.029$        & $-0.152 \pm 0.018$        & $-0.077 \pm 0.011$    \\
    2           & $-0.206 \pm 0.130$        & $-0.221 \pm 0.089$        & $-0.103 \pm 0.058$        & $-0.019 \pm 0.034$        & $-0.005 \pm 0.020$        & $-0.011 \pm 0.013$    \\
    3           & $-0.352 \pm 0.106$        & $-0.383 \pm 0.075$        & $-0.331 \pm 0.052$        & $-0.185 \pm 0.031$        & $-0.082 \pm 0.018$        & $-0.032 \pm 0.012$    \\
    \hline
    Combined    & $-0.371 \pm 0.065$        & $-0.412 \pm 0.046$        & $-0.288 \pm 0.031$        & $-0.154 \pm 0.018$        & $-0.080 \pm 0.011$        & $-0.040 \pm 0.007$    \\
    \hline
    \end{tabular*}
    \caption{Calculated absorption depth (\%) of the Na doublet for each transmission spectrum following the airmass (a) telluric removal method. The flux is averaged across six passbands with total widths of 0.375, 0.75, 1.5, 3, 6, and 12~{\AA}. Combined depths are measured after co-adding individual nightly spectra.}
    \label{tab:absorption_depths_a}
\end{table*}

\begin{table*}
    \centering
    \begin{tabular*}{\textwidth}{c@{\extracolsep{\fill}}cccccc}
    \hline
    Night       & AD$^{0.375}_{\text{m}}$   & AD$^{0.75}_{\text{m}}$    & AD$^{1.5}_{\text{m}}$     & AD$^{3}_{\text{m}}$       & AD$^{6}_{\text{m}}$       & AD$^{12}_{\text{m}}$  \\
    \hline
    1           & $-0.430 \pm 0.101$        & $-0.525 \pm 0.072$        & $-0.342 \pm 0.049$        & $-0.183 \pm 0.029$        & $-0.102 \pm 0.017$        & $-0.050 \pm 0.011$    \\
    2           & $-0.585 \pm 0.132$        & $-0.553 \pm 0.091$        & $-0.353 \pm 0.059$        & $-0.180 \pm 0.035$        & $-0.089 \pm 0.021$        & $-0.063 \pm 0.013$    \\
    3           & $-0.398 \pm 0.107$        & $-0.425 \pm 0.076$        & $-0.355 \pm 0.052$        & $-0.206 \pm 0.031$        & $-0.095 \pm 0.018$        & $-0.034 \pm 0.012$    \\
    \hline
    Combined    & $-0.471 \pm 0.066$        & $-0.501 \pm 0.046$        & $-0.350 \pm 0.031$        & $-0.190 \pm 0.018$        & $-0.095 \pm 0.011$        & $-0.049 \pm 0.007$    \\
    \hline
    \end{tabular*}
    \caption{Same as Table~\ref{tab:absorption_depths_a} but measuring from the transmission spectrum after removing telluric contamination with \texttt{molecfit} (m).}
    \label{tab:absorption_depths_m}
\end{table*}
\endgroup

\subsection{Binned sodium absorption}
\label{sec:results_depths}

HD~189733~b has been studied extensively in the past at different resolutions \citep{redfield2008, jensen2011, huitson2012, wyttenbach2015, khalafinejad2017}. In order to form a comparison to these (and studies of other planets), it is beneficial to calculate the strength of the absorption over varying bin widths centred at the sodium lines \citep{snellen2008, wyttenbach2015}. This is particularly useful if the line itself is too narrow to be resolved, or if there is too much noise to produce an accurate fit.

We define absorption depth (AD) as the difference between the mean flux of the continuum regions and the mean flux within a passband centred on the spectral feature. Red and blue continuum passbands are chosen in regions around the sodium doublet: 5874.89--5886.89~{\AA} for the blue band, and 5898.89--5910.89~{\AA}~for the red band. The mean flux in central passbands of widths 0.375, 0.75, 1.5, 3, 6, and 12~{\AA} are measured. Since the sodium feature contains two lines, the total widths of these passbands are divided into two, with one half centred around each line. However, the 12~{\AA} passband is large enough to encapsulate both features, so this is not split and is centred in the middle of the two lines instead. The absorption depth is thus calculated by 
\begin{equation}
    \text{AD} = \overline{\Re'_{\text{C}}} - \frac{\overline{\Re'_{\text{B}}} + \overline{\Re'_{\text{R}}}}{2} ~,
    \label{eqn:passband}
\end{equation}
where $\overline{\Re'}$ is the mean flux of the transmission spectrum in the blue (B), red (R), and central (C) passbands. This quantity was evaluated over all passbands for each nightly and combined transmission spectrum. All averages were weighted using the inverse of the squared uncertainties (variance), which were propagated from the photon noise of the raw data. The results are shown in Table~\ref{tab:absorption_depths_a} (airmass) and Table~\ref{tab:absorption_depths_m} (\texttt{molecfit}). The measured absorption depths of the combined data are shallower for the airmass method than the \texttt{molecfit} method. This matches the difference seen in the line contrasts in section~\ref{sec:results_telluric_removal}.

Narrower passbands more accurately probe the sodium line cores, so these measurements will most closely match the Gaussian line contrasts. For the \texttt{molecfit} reduction, the absorption depths were $-0.471 \pm 0.066~\%$ and $-0.501 \pm 0.046~\%$ for the 0.375 and 0.75~{\AA} passbands respectively. These values are consistent to within 1$\sigma$ of each other, and the error for the 0.75~{\AA} passband overlaps with the error in the average Gaussian line contrast. The same comparison is true for the airmass method.

We compare our results to those of \citet{wyttenbach2015}, who analyse the same HARPS data and compare the effects of planetary radial velocity corrections. In the 1.5~{\AA} passband, they measure an absorption depth of $-0.320 \pm 0.031~\%$, which agrees to within 1$\sigma$ with our results of $-0.288 \pm 0.031~\%$ for the airmass method and $-0.350 \pm 0.031~\%$ for the \texttt{molecfit} method. We therefore confirm the 10$\sigma$ detection of sodium in the atmosphere of HD~189733~b.

Table~\ref{tab:absorption_depths_a} shows slight variation to the results discussed by \citet{wyttenbach2015}, despite similarities in the telluric removal methods. This is likely due to differences in the reduction pipeline and handling of the data, as well as corrections for the CLV and RM effects which change the shape and depth of the absorption lines. A consistent trend is seen in both studies: the measured depths for night 2 using the airmass reduction are significantly lower than the other two nights. As discussed previously, the observations on this night began when the planet was already in-transit, so a long out-of-transit baseline was not obtained. A clear telluric reference spectrum was only able to be derived when the in-transit spectra were included. Whilst this is better for telluric correction, it could also lead to partial removal of signals in the transmission spectrum. We were unable to detect any significant absorption in the D1 line for this night, which gives rise to the lower overall measurements. The \texttt{molecfit} reduction performs better in this case. As seen in Table~\ref{tab:absorption_depths_m}, the measured depths are much more consistent across all nights for each passband, and there is a significant improvement in the results for night 2. Therefore, \texttt{molecfit} is much better at removing telluric contamination in spectra, regardless of the number of in or out-of-transit observations. It is also noted that the errors across single passbands are comparable for both methods, therefore the signal-to-noise ratio is improved when using \texttt{molecfit}.

As the passbands increase in size, the measured depths for both telluric correction methods decrease by the same proportion. We measure an absorption depth of $-0.049 \pm 0.007~\%$ in the largest (12~{\AA}) passband using \texttt{molecfit}. This agrees with the space-based detection of $-0.051 \pm 0.006~\%$ \citep{huitson2012}, and two other ground-based detections of $-0.067 \pm 0.020~\%$ \citep{redfield2008} and $-0.053 \pm 0.017~\%$ \citep{jensen2011} within the same 12~{\AA} passband. The corresponding measurement for the airmass method was $-0.040 \pm 0.008~\%$, which does not agree as significantly. This suggests that \texttt{molecfit} is better at correcting tellurics in the immediate regions around the sodium doublet.

\subsection{Wind speeds}
\label{res:winds}

Sodium absorption features in the lab rest frame should be located at wavelengths of 5889.951~{\AA} for the D2 line and 5895.924~{\AA} for the D1 line. Despite efforts made to correct for spectral deviations due to systemic and planetary radial velocities, the measured centroids of the Gaussian fits to the full-resolution data (Table~\ref{tab:gauss_fit}) have a net blueshift from the rest frame Na doublet positions (although still encapsulated within error). We interpret this as being due to winds within the atmosphere of HD~189733~b, in agreement with other studies \citep{wyttenbach2015, louden2015, casasayas-barris2017}. The average blueshift was $-0.025 \pm 0.029$~{\AA} for the airmass method and $-0.035 \pm 0.025$~{\AA} for the \texttt{molecfit} method. This corresponds to net wind velocities of $-1.3 \pm 1.5$~km~s$^{-1}$ and $-1.8 \pm 1.2$~km~s$^{-1}$ respectively, indicating an eastward atmospheric movement from day-side to night-side.

Since the telluric corrections are the only parts that vary between our methods, we deduce that the difference between the velocities may be due to inaccuracies in telluric removal -- particularly in nights 2 and 3 for the airmass reduction. Nevertheless, both results are lower than the $-8 \pm 2$~km~s$^{-1}$ measurement discussed by \citet{wyttenbach2015}. The discrepancy can be attributed to nuances in the reduction pipeline and radial velocity corrections, and application of models to correct for CLV and RM effects. If the spectra are observed uniformly over the full transit and stacked in the stellar rest frame, the RM effect can be averaged out. As shown in Figure~\ref{fig:rm_correction}, this is not true when the transmission spectra are stacked in the planetary rest frame -- spurious signals are induced (albeit small for the slowly rotating host star) and were removed by modelling these effects.

Our measured wind velocities are consistent with the results of \citet{louden2015} ($-1.9^{+0.7}_{-0.6}$~km~s$^{-1}$) and \citet{casasayas-barris2017} ($-2$~km~s$^{-1}$) who analysed the same data. Additionally, \citet{louden2015} spatially resolved the atmospheric winds and measured a redshift on the leading limb of the planet ($2.3^{+1.3}_{-1.5}$~km~s$^{-1}$) and a blueshift on the trailing limb ($-5.3^{+1.0}_{-1.4}$~km~s$^{-1}$). This implies a net eastward motion due to a combination of tidally locked planetary rotation and strong eastward winds. High-resolution observations in the infrared regime have also reported similar day-side to night-side wind speeds: $-1.7^{+1.1}_{-1.2}$~km~s$^{-1}$ \citep{brogi2016} and $-1.6^{+3.2}_{-2.7}$~km~s$^{-1}$ \citep{brogi2018}. Additionally, our measured wind velocities are consistent with predictions from three-dimensional General Circulation Models (GCMs), where the atmospheric flow is dominated by a super-rotating eastward equatorial jet and the hottest region of the atmosphere is advected downwind from the substellar point \citep{showman2009, kempton2012, rauscher2014}. For HD~189733~b, models predict that the equatorial jet extends to latitudes of ${\sim}60^{\circ}$. Atmospheric winds flow eastward within this region -- from night-to-day at the western terminator (leading limb) and from day-to-night at the eastern terminator (trailing limb). At latitudes above ${\sim}60^{\circ}$, the winds flow preferentially from day-to-night at both the leading and trailing limbs of the terminator. Therefore, models predict a net blueshift due to the winds \citep{showman2013}.

\subsection{Extent of sodium in the atmosphere}

An exoplanet's transit depth is wavelength-dependent; if there is molecular or atomic absorption, the atmosphere will appear opaque when viewed at wavelengths within this region. The planet will have a larger apparent radius at these wavelengths and block more stellar flux, compared to a transparent atmosphere with no absorption. Therefore, we can use absorption measurements to probe the atmospheric height at which the molecules or atoms are present. We define atmospheric scale height $H_{\text{sc}}$ as the increase in altitude over which the pressure reduces by a factor of $e$, given by
\begin{equation}
    H_{\text{sc}} = \frac{k_{\text{B}}T_{\text{eq}}}{\mu_{\text{m}}g} ~,
    \label{eqn:scale_height}
\end{equation}
where $k_{\text{B}}$ is the Boltzmann constant, $T_{\text{eq}}$ is the equilibrium temperature, $\mu_{\text{m}}$ is the mean molecular weight, and $g$ is the planet's surface gravity.
Using $\mu_{\text{m}} = 2.22$~u (for Jupiter) and deriving $g = 22.7$~m s$^{-2}$ from the parameters in Table~\ref{tab:parameters}, the scale height for HD~189733~b is 200.7~km. We refer to the white-light transit depth as the decrease in stellar flux during transit, equal to the ratio of the sky-projected areas of the planet and star:
\begin{equation}
    \Delta_0 = \left(\frac{R_{\text{p}}}{R_\ast}\right)^2 ~.
    \label{eqn:white_light_transit}
\end{equation}
For HD~189733~b, $\Delta_0 = 0.0228$. Atmospheric absorption increases the apparent planetary radius by a height $H_\lambda$, so a wavelength-dependent transit depth is similarly defined as
\begin{equation}
    \Delta_\lambda = \left(\frac{R_{\text{p}} + H_\lambda}{R_\ast}\right)^2 = \left(\frac{R_{\text{p}}}{R_\ast}\right)^2 + \frac{2R_{\text{p}}H_\lambda}{R_\ast^2} + \left(\frac{H_\lambda}{R_\ast}\right)^2 ~.
    \label{eqn:wavelength_transit}
\end{equation}
Since the atmosphere typically extends around 5--10 scale heights \citep{madhusudhan2019}, $H_\lambda << R_\ast$, so 
\begin{equation}
    \Delta_\lambda = \Delta_0 + \frac{2R_{\text{p}}H_\lambda}{R_\ast^2} ~.
    \label{eqn:wavelength_transit_2}
\end{equation}
For a particular wavelength, the difference between the white-light and wavelength-dependent transit depths is a measure of the absorption,
\begin{equation}
    \delta_\lambda = \Delta_\lambda - \Delta_0 = \frac{2R_{\text{p}}H_\lambda}{R_\ast^2} = \frac{2\Delta_{0}H_\lambda}{R_{\text{p}}} ~,
    \label{eqn:absorption_depth}
\end{equation}
where
\begin{equation}
    \delta_\lambda = 1 - \frac{F_{\lambda, \text{in}}}{F_{\lambda, \text{out}}} = -{\Re'}_{\lambda} ~.
    \label{eqn:absorption_depth_Re}
\end{equation}
Using the line contrasts and passband absorption depths as a measurement of ${\Re'}_{\lambda}$, we can therefore calculate the height of an atmospheric detection by rearranging equation~\ref{eqn:absorption_depth} for $H_\lambda$. For the \texttt{molecfit} reduction, the average Na doublet line contrast of $-0.59 \pm 0.05~\%$ corresponds to an apparent planetary radius of $1.13~R_{\text{p}}$ and an atmospheric height of $H\textsubscript{Na} = 10200 \pm 900$~km. This is ${\sim}51$ times greater than the scale height and suggests that the detection of Na probes the upper atmosphere of HD~189733~b.

\section{Summary and Conclusion}
\label{sec:conclusion}

HD~189733~b is one of the most extensively studied exoplanets to date. A number of chemical species have been detected through observations from space and ground-based instruments, as well as measurements of wind speeds, atmospheric circulation, and evidence of Rayleigh scattering and high-altitude hazes. The nature of telluric contamination makes ground-based observations of exoplanet transmission spectra difficult. Several methods have been used in the past to remove excess absorption from Earth's atmosphere, including observing a telluric standard star \citep{redfield2008, jensen2011, jensen2012}, computing an empirical telluric spectrum \citep{wyttenbach2015}, and using an Earth atmospheric model \citep{yan2015, casasayas-barris2017, allart2017}.

In this study we approached the challenge of removing telluric lines in high-resolution optical spectra by comparing two popular methods using archival HARPS observations of HD~189733~b. First, we derived a telluric reference spectrum from the data by assuming a linear relationship between airmass and telluric line strength. The reference spectrum was used to scale the strength of telluric lines to a constant level as if they had been observed at the same airmass, and are thus removed during the subsequent transmission spectrum calculation. Our second method involved using \texttt{molecfit} to generate a model of molecular H$_2$O and O$_2$ absorption in Earth's atmosphere. It is then straightforward to remove the contamination through division of unique telluric contributions for each observation. The methods can be applied to all regions of the HARPS spectra, however we focused specifically on the quality of the telluric correction around the sodium D-lines and the impact on the measured line properties. Other empirical methods such as \texttt{SYSREM} or principal component analysis \citep{tamuz2005, mazeh2007, birkby2013}, which are normally reserved for severe telluric contamination in the near-infrared, are beyond the scope of this study.

Alongside correcting for stellar, systemic and planetary radial velocity effects, we modelled the Centre-to-Limb Variation and Rossiter-McLaughlin effect to correct for spurious signals which are imprinted onto the transmission spectrum. The relative strengths of these signals were small and at the noise level of the spectrum, however it is an important factor to consider if the star is rotating more rapidly. Both methods were performed on the data from each night separately, then combined for an overall transmission spectrum.

Absorption due to atmospheric sodium (Na \textsc{i}) was identified from the doublet lines at ${\sim}5890$~{\AA} (D2) and ${\sim}5896$~{\AA} (D1). For the airmass reduction, we measured Gaussian line contrasts of $-0.55 \pm 0.07~\%$ (D2) and $-0.45 \pm 0.07~\%$ (D1), and absorption in the 0.75~{\AA} passband of $-0.412 \pm 0.046~\%$. For the \texttt{molecfit} reduction, the corresponding values were $-0.64 \pm 0.07~\%$ (D2 line contrast), $-0.53 \pm 0.07~\%$ (D1 line contrast), and $-0.501 \pm 0.046~\%$ (0.75~{\AA} passband absorption). In cases where weather conditions are variable over one night, the model produced with \texttt{molecfit} was more robust at removing telluric contamination, and absorption depth measurements were more consistent across all nights. Additionally, it performs better when the observing window does not allow for a long out-of-transit baseline to be obtained (when building an empirical telluric reference spectrum is difficult). The average line contrast of the Na doublet for the \texttt{molecfit} method was $-0.59 \pm 0.05~\%$, corresponding to an atmospheric height of $10200 \pm 900$~km. Finally, we measured a net atmospheric blueshift of $-1.8 \pm 1.2$~km~s$^{-1}$ arising from a combination of tidally locked planetary rotation and day-side to night-side winds.

From the assessments of the results, we conclude that correcting telluric contamination with a model of molecular (H$_2$O and O$_2$) absorption in Earth's atmosphere using \texttt{molecfit} produces a better quality transmission spectrum than if the tellurics were removed empirically, e.g. using airmass. This is evident by: (1) a detection of atmospheric sodium with greater significance, (2) more consistent measurements of absorption depths in bandpasses across different nights, and (3) the ability to extract a signal when the weather conditions are not ideal. \texttt{Molecfit} allows for targeted removal of specific telluric lines without significantly impacting data outside of these regions. This is a key advantage over data-driven methods such as the airmass correction used within this work, since it is likely that the empirically derived model will more significantly increase the continuum noise. Additionally, corrections can be performed on a single spectrum without relying on a large sample of observations, which is particularly advantageous if only a partial transit is obtained or if variable weather conditions affect a large number of spectra.

Detections at different spectral resolutions and across a number of wavelength regimes are important for thorough atmospheric characterisation. With an ever-growing number of identifications of chemical species within exoplanet atmospheres, we are able to further understand and constrain dynamical and chemical process and make inferences into how hot Jupiters form and evolve. We have already seen how ground-based instruments such as HARPS can be a powerful tool for such studies in the optical regime, and this is set to increase with newly developed high-resolution spectrographs such as ESPRESSO \citep{pepe2013}, HARPS-3 \citep{young2018}, and EXPRES \citep{jurgenson2016}. ESPRESSO covers a wavelength range of 380--780~nm which probes deeper into the red wavelengths, giving opportunity to make detections of the water band at ${\sim}7400$~{\AA} or the potassium doublet at ${\sim}7665$--7699~{\AA}. It is therefore critical to be able to tackle the problem of telluric contamination in a robust way across a broad wavelength range to ensure accurate recovery of the planetary transmission spectrum.

\section*{Acknowledgements}
We thank the anonymous referee for their thoughtful comments which helped to improve the quality of this manuscript. AL acknowledges support from the Science and Technology Facilities Council (STFC), UK. AL thanks Annelies Mortier for insightful discussion about the HARPS spectrograph and data.

\section*{Data Availability}
All the data used in this work are available through the ESO science archive facility. The data are based on observations collected at the European Southern Observatory under ESO programmes  072.C-0488(E), 079.C-0828(A) (PI: Mayor) and 079.C-0127(A) (PI: Lecavelier des Etangs).


\bibliographystyle{mnras}
\bibliography{ms} 


\appendix

\section{System Parameters}

\begingroup
\renewcommand{\arraystretch}{1.5} 
\begin{table*}
    \centering
    \begin{tabular*}{\textwidth}{l@{\extracolsep{\fill}}llll}
    \hline
    Parameter                       & Symbol        & Value                         & Unit                      & Reference     \\
    \hline
    \hline
    \textbf{Star}                   &               &                               &                           &               \\
    \hline
    Stellar Mass                    & $M_\ast$     & 0.823 $^{+0.022}_{-0.029}$     & M$_{\sun}$                & \citet{triaud2009} \\
    Stellar Radius                  & $R_\ast$     & 0.756 $^{+0.018}_{-0.018}$     & R$_{\sun}$                & \citet{torres2008} \\
    Stellar RV Semi-amplitude       & $K_\ast$     & 200.56 $^{+0.88}_{-0.88}$      & m s$^{-1}$                & \citet{boisse2009} \\
    Effective Temperature           & $T_{\text{eff}}$     & 5052 $^{+16}_{-16}$    & K                         & \citet{stassun2017} \\
    Projected Rotational Velocity   &$v\sin{i}$     & 3.5 $^{+1.0}_{-1.0}$          & km s$^{-1}$               & \citet{bonomo2017} \\
    Surface Gravity                 & $\log{g}$     & 4.49 $^{+0.05}_{-0.05}$       & $\log_{10}$(cm s$^{-2}$)  & \citet{stassun2017} \\
    Metallicity                     & [Fe/H]        & -0.03 $^{+0.08}_{-0.08}$      & dex                       & \citet{torres2008} \\
    \hline
    \textbf{Planet}                 &               &                               &                           &               \\
    \hline
    Planetary Mass                  & $M_{\text{p}}$ & 1.138 $^{+0.022}_{-0.025}$   & M$_{\text{J}}$            & \citet{triaud2009} \\
    Planetary Radius                & $R_{\text{p}}$ & 1.138 $^{+0.027}_{-0.027}$   & R$_{\text{J}}$            & \citet{torres2008} \\
    Planetary RV Semi-amplitude     & $K_{\text{p}}$ & 151 $^{+6}_{-6}$             & km s$^{-1}$               & Derived      \\
    Equilibrium Temperature         & $T_{\text{eq}}$ & 1220 $^{+13}_{-13}$         & K                         & \citet{addison2019} \\
    Orbital Inclination             & $i_{\text{p}}$ & 85.710 $^{+0.024}_{-0.024}$  & deg                       & \citet{agol2010} \\
    Sky Projected Obliquity         & $\lambda$     & -0.4 $^{+0.2}_{-0.2}$         & deg                       & \citet{cegla2016} \\
    \hline
    \textbf{System}                 &               &                               &                           &                \\
    \hline
    Period                          & $P$    & 2.21857567 $^{+0.00000015}_{-0.00000015}$   & days               & \citet{agol2010} \\
    Mid-transit Time                & $T_0$  & 2454279.436714 $^{+0.000015}_{-0.000015}$   & BJD\textsubscript{UTC} & \citet{agol2010}\\
    Transit Duration                & $\tau$        & 0.07527 $^{+0.00020}_{-0.00037}$     & days               & \citet{triaud2009} \\
    Semi-major axis                 & $a$           & 0.03120 $^{+0.00027}_{-0.00037}$     & A.U.               & \citet{triaud2009} \\
    Systemic Velocity               & $v_{\text{sys}}$ & 2.2765 $^{+0.0017}_{-0.0017}$     & km s$^{-1}$        & \citet{boisse2009} \\
    \hline
    \end{tabular*}
    \caption{Stellar, planetary, and orbital parameters of the HD~189733 system. $K_{\text{p}}$ is derived through the relationship $-K_{\ast}(M_{\ast}/M_{\text{p}})$.}
    \label{tab:parameters}
\end{table*}
\endgroup


\bsp	
\label{lastpage}
\end{document}